\documentclass[12pt]{article}
\pdfoutput=1

\usepackage{putex}
\usepackage{graphicx}
\usepackage{caption}
\usepackage{amsmath}
\usepackage{array}
\usepackage{subcaption}
\usepackage{epstopdf}
\usepackage{enumerate}
\usepackage{cite}
\usepackage{youngtab}
\usepackage{tensor}
\usepackage{slashed}
\usepackage[aligntableaux=center]{ytableau}
\usepackage[utf8]{inputenc}
\usepackage{rotating}
\usepackage{bigfoot}
\usepackage[
      colorlinks=true,
      linkcolor=blue,
      urlcolor=blue,
      filecolor=black,
      citecolor=red,
      ]{hyperref}

\newcommand {\be} {\begin {equation}}
\newcommand {\ee} {\end {equation}}

\newcommand {\bes} {\begin {equation*}}
\newcommand {\ees} {\end {equation*}}

\newsavebox{\numberedequationbox}
\newcommand{\es}[2]{%
\begin{equation}\label{#1}%
\sbox{\numberedequationbox}{$\displaystyle\begin{aligned}#2\end{aligned}$}%
\ifdim\wd\numberedequationbox>.90\linewidth
\resizebox{.90\linewidth}{!}{\usebox{\numberedequationbox}}%
\else\usebox{\numberedequationbox}\fi
\end{equation}}

\newcommand{\cA}{{\mathcal A}}

\newcommand{\cG}{{\mathcal G}}

\newcommand{\cS}{{\mathcal S}}

\newcommand{\beq}{\begin{equation}}
\newcommand{\eeq}{\end{equation}}

\def\ie{\begin{equation}\begin{aligned}}
\def\fe{\end{aligned}\end{equation}}

\numberwithin{equation}{section}

\def\<{\langle}
\def\>{\rangle}

\begin{document}

\preprint{}

\institution{oxford}{Mathematical Institute, University of Oxford,
Woodstock Road, Oxford, OX2 6GG, UK}
\institution{Exile}{Department of Particle Physics and Astrophysics, Weizmann Institute of Science, Rehovot, Israel}
\institution{paris}{\unexpanded{\shortstack{Universit\'e Paris-Saclay, CNRS, CEA, Institut de Physique Th\'eorique, \\91191, Gif-sur-Yvette, France}}}

\title{M-theory on $AdS_4\times S^7$ at 1-loop and beyond}

\authors{Luis F. Alday\worksat{\oxford}, Shai M. Chester\worksat{\Exile}, and Himanshu Raj\worksat{\paris}}

\abstract{
We study graviton scattering on $AdS_4\times S^7$, which is dual to the stress tensor multiplet four-point function in the maximally supersymmetric 3d $U(N)_1\times U(N)_{-1}$ ABJM theory. We compute 1-loop corrections to this holographic correlator coming from Witten diagrams with supergravity $R$ and higher derivative $R^4$ vertices, up to contact term ambiguities, and find that the flat space limit matches the corresponding terms in the 11d M-theory S-matrix. We then use supersymmetric localization to show that  all the 1-loop contact terms vanish, as was previously observed for the  $AdS_4\times S^7/\mathbb{Z}_2$ theory dual to $U(N)_2\times U(N)_{-2}$ ABJM. Finally, we use the recent localization results of Gaiotto and Abajian, as inspired by twisted M-theory, to compute all the short OPE coefficients in correlators of the stress tensor multiplet and the next lowest half-BPS operator, which we find saturate the bootstrap bounds on these mixed correlators for all $N$.
}
\date{}

\maketitle

\tableofcontents

\section{Introduction}
\label{intro}

M-theory is harder to study than string theory. The reason string theory is easier is due to the worldsheet description, which can be used to compute observables like the graviton S-matrix at finite Planck length in a small string coupling expansion. For M-theory, however, there is no such description, and very little is known about the S-matrix. In particular, we can expand the M-theory S-matrix at small Planck length $\ell_{11}$ as
\es{A}{
\mathcal{A}(s,t)=\ell_{11}^{9}\mathcal{A}_{R}+\ell^{15}_{11}\mathcal{A}_{R^4}+\ell^{18}_{11}\mathcal{A}_{R|R}+\ell^{21}_{11}\mathcal{A}_{D^6R^4}+\ell^{23}_{11}\mathcal{A}_{D^8R^4}+\ell_{11}^{24}\mathcal{A}_{R|R^4}+\dots\,,
}
where $s,t,u$ are 11d Mandelstam variables. The lowest few terms involve the protected vertices $R$, $R^4$, and $D^6R^4$, which have been derived in various papers \cite{Green:1997as, Russo:1997mk, Green:2005ba,Alday:2020tgi} by using the duality to Type IIA string theory when the eleventh direction is small.\footnote{See \cite{Duff:1995wd} for discussion of the $R^4$ correction from the perspective of anomaly cancellation in 11d.} Terms involving unprotected vertices, such as the lowest unprotected term $D^8R^4$ shown above, cannot be fixed using the duality to Type IIA, and so are still unknown. 

In \cite{Chester:2018aca}, a new strategy to study the M-theory S-matrix was introduced, which uses the holographic duality between M-theory on $AdS_4\times S^7/\mathbb{Z}_k$ and $U(N)_k\times U(N)_{-k}$ ABJM theory. In this duality, scattering of gravitons and other KK modes in the bulk are related to correlators of single trace operators such as the stress tensor multiplet in the CFT, which are in principle easier to study. The flat space S-matrix can then be obtained by a precise flat space limit applied to the CFT correlator \cite{Penedones:2010ue}.

In this paper, we will study these holographic correlators using two complementary approaches. Firstly, we consider the large $c_T$ expansion of the stress tensor correlator $\langle2222\rangle$, which is related to the small $\ell_{11}$ expansion according to the dictionary \cite{Aharony:2008ug,Aharony:2008gk}:
\es{cPlanck}{
  \frac{L^6}{\ell_{11}^6}=\left(\frac{3\pi c_T k}{2^{11}}\right)^{\frac23}+O(c_T^0) \,,
}
where the stress tensor two-point function coefficient $c_T$ is related to $N$ as $c_T\sim N^{3/2}$. The stress tensor correlator is most easily expressed in Mellin space \cite{Fitzpatrick:2011ia}, where the analytic bootstrap \cite{Rastelli:2017udc} constrains its large $c_T$ expansion to take the form
\es{M2222}{
M(s,t;\sigma,\tau)&=c_T^{-1}M^R+c_T^{-\frac53}B^{R^4}_4M^4+c_T^{-2}(M^{R|R}+B^{R|R}_4M^4)\\
&+c_T^{-\frac73}(B^{D^6R^4}_4M^4+B^{D^6R^4}_6M^6+B^{D^6R^4}_7M^7)\\
&+c_T^{-\frac{23}{9}}(B^{D^8R^4}_4M^4+B^{D^8R^4}_6M^6+B^{D^8R^4}_7M^7+B^{D^8R^4}_8M^8)\\
&+c_T^{-\frac83}(M^{R|R^4}+B^{R|R^4}_4M^4+B^{R|R^4}_6M^6+B^{R|R^4}_7M^7+B^{R|R^4}_8M^8)+\dots\,,
}
where the $M$'s are functions of $s,t,\sigma,\tau$ with numerical coefficients $B$ that can depend on $k$, and $M^a(s,t)$ are polynomials in $s,t$ of degree $a$. The tree level terms\footnote{Since $c_T$ is the only expansion parameter, we can only distinguish between tree and loop terms at low orders where they have different power of $c_T$.} at orders $c_T^{-1}$, $c_T^{-\frac53}$, and $c_T^{-\frac73}$ were previously computed for both $k=1,2$ ABJ(M) in \cite{Zhou:2017zaw}, \cite{Chester:2018aca}, and  \cite{Binder:2018yvd}, respectively. The 1-loop terms $R|R$ at order $c_T^{-2}$ and $R|R^4$ at order $c_T^{-\frac83}$ were computed in \cite{Alday:2021ymb} for $k=2$, where it was found that the supersymmetric localization constraints of \cite{Chester:2018aca,Binder:2018yvd} fixed the coefficients $B_i$ of the contact terms to zero, and the flat space limit matched the corresponding terms in the 11d S-matrix. 

Here, we will generalize this calculation to $k=1$, and find similar results. The 1-loop amplitude is computed following \cite{Aharony:2016dwx} by fixing the double discontinuity at 1-loop in terms of tree level data, which can then be used to extract all CFT data using the Lorentzian inversion formula \cite{Caron-Huot:2017vep}, and can also be used to fix the physical contact term ambiguities $B_i$. While this data captures in principle all the physical content of the 1-loop amplitude, it is sometimes useful to write down an explicit Mellin space expression, which we do in the attached \texttt{Mathematica} file in terms of various ambiguities that are all in principle fixed by superconformal symmetry, but are difficult to compute in practice for $R|R$.\footnote{The explicit $R|R$ Mellin amplitude for $k=2$ was presented in the same form in \cite{Alday:2021ymb}} The main technical difficulty relative to previous 1-loop calculations such as the $k=2$ theory and results in 4d \cite{Alday:2017xua,Alday:2018pdi,Alday:2018kkw,Alday:2019nin,Alday:2017vkk,Aprile:2017bgs,Aprile:2017qoy,Aprile:2019rep,Drummond:2019hel,Drummond:2020dwr,Aprile:2020luw} and 6d \cite{Alday:2020tgi}, is that the double discontinuity in the $k=1$ case also receives contributions from tree level data of OPE coefficients with odd twists,\footnote{This was first observed for non-supersymmetric 3d correlators with weakly broken higher spin symmetry in \cite{Aharony:2018npf}.} which are much harder to compute.

We will also study the holographic correlator at finite $c_T$ using the numerical bootstrap combined with constraints from supersymmetric localization. The numerical bootstrap can be used to compute bounds on CFT data in $\langle2222\rangle$ that apply to any maximally supersymmetric 3d CFT. In \cite{Alday:2022ldo}, it was observed that these most general bounds are saturated by the holographic dual to pure AdS$_4$ supergravity, whose only single trace operator is the stress tensor multiplet, so if we want bounds that are related to M-theory we need to either input additional information about the CFT or consider mixed correlator bounds with other single trace operators. The former strategy was pursued for the $k=2$ theory in \cite{Alday:2021ymb}, where it was found that inputting the values of certain short OPE coefficients, which can be computed using supersymmetric localization to all orders in $1/N$ \cite{Agmon:2017xes}, strengthens the bootstrap bounds such that they are then saturated by the 1-loop correction to M-theory on $AdS_4\times S^7/\mathbb{Z}_2$. 

For the $k=1$ theory we consider here, we would like to consider mixed correlators of the stress tensor multiplet and the next lowest half-BPS operator, since this setup automatically excludes the pure AdS$_4$ theory as well as the $k=2$ theory, which do not contain this other half-BPS operator. This mixed correlator setup was considered in \cite{Agmon:2019imm}, but the full set of short operator OPE coefficients was not known to all orders in $1/N$.\footnote{Supersymmetric localization was used to compute these OPE coefficients in terms of $N^2$ integrals \cite{Agmon:2017lga,Dedushenko:2016jxl,Agmon:2019imm}, but these are hard to compute at large $N$.} Here, we use the recent results of Gaiotto and Abajian \cite{Gaiotto:2020vqj}, which were inspired by twisted M-theory \cite{Costello:2016nkh,Gaiotto:2019wcc}, to compute these OPE coefficients to all orders in $1/N$. We find that after inputting a subset of these coefficients, the bootstrap bounds for the remaining independent coefficients are saturated by their all orders in $1/N$ values, which suggests that supplementing bootstrap with localization is sufficient to fix the physical theory.

 The rest of this paper is organized as follows. In Section \ref{1loop} we compute the 1-loop correction to $\langle2222\rangle$, and check that the flat space limit matches the 11d S-matrix. In Section \ref{protOPE}, we use supersymmetric localization to fix the contact terms in this correlator, and to compute all short OPE coefficients that appear in the mixed correlator system. In Section \ref{numBoot} we compare these localization results to bounds from the numerical conformal bootstrap. We end with a conclusion in Section \ref{conc}, where we summarize our results and discuss future directions. We also include an ancillary \texttt{Mathematica} notebook, which includes many of our more complicated explicit results.

\section{$\langle2222\rangle$ at 1-loop}
\label{1loop}

In this section we will compute the $\langle2222\rangle$ holographic correlator at 1-loop in the $1/c_T$ expansion. We start by reviewing the superblock expansion for the $\langle22pp\rangle$ correlator. We then discuss how to compute the tree level $\langle22pp\rangle$ CFT data that we will need. Finally, we show how to use this data, along with generalized free field theory (GFFT) results from \cite{Alday:2021ymb}, to compute the 1-loop double discontinuities for $\langle2222\rangle$, which can be used to extract their 1-loop CFT data with sufficiently large spin.

\subsection{Setup}
\label{setup}

We consider half-BPS superconformal primaries $S_p$ in 3d $\mathcal{N}=8$ SCFTs that are scalars with $\Delta=\frac p2$ and transform in the $[00p0]$ of $\mathfrak{so}(8)_R$,\footnote{The convention we use in defining these multiplets is that the supercharges transform in the ${\bf 8}_v = [1000]$ irrep of $\mathfrak{so}(8)_R$.} where $p=1,2,\dots$. The first such interacting operator is $S_2$, which is the bottom component of the stress tensor multiplet. We can denote these operators as traceless symmetric tensors $S_{I_1\dots I_p}( x)$ of $\mathfrak{so}(8)_R$, where $I_i=1,\dots8$. We can avoid indices by introducing an auxiliary polarization vector $Y^I$ that is constrained to be null, $Y_i\cdot Y_i=0$, and then define
\es{S}{
S_p( x,Y)\equiv S_{I_1\dots I_p}Y^{I_1}\cdots Y^{I_p}\,.
}

Consider the four-point functions $\langle 22pp\rangle$. Conformal and $\mathfrak{so}(8)_R$ symmetry fixes these correlators to take the form
\es{4point}{
\langle S_2( x_1,Y_1)S_2( x_2,Y_2)S_p( x_3,Y_3)S_p( x_4,Y_4) \rangle=\frac{(Y_1\cdot Y_2)^2(Y_3\cdot Y_4)^p}{| x_{12}|^{2}| x_{34}|^{p}}\mathcal{G}_{p}(U,V;\sigma,\tau)\,,
}
where we define
 \es{uvsigmatauDefs}{
  U \equiv \frac{{x}_{12}^2 {x}_{34}^2}{{x}_{13}^2 {x}_{24}^2} \,, \qquad
   V \equiv \frac{{x}_{14}^2 {x}_{23}^2}{{x}_{13}^2 {x}_{24}^2}  \,, \qquad
   \sigma\equiv\frac{(Y_1\cdot Y_3)(Y_2\cdot Y_4)}{(Y_1\cdot Y_2)(Y_3\cdot Y_4)}\,,\qquad \tau\equiv\frac{(Y_1\cdot Y_4)(Y_2\cdot Y_3)}{(Y_1\cdot Y_2)(Y_3\cdot Y_4)} \,,
 }
 with $x_{ij}\equiv x_i-x_j$. We will sometimes drop the $2$ subscript for the stress tensor correlator as well as its Mellin transform, e.g. \eqref{M2222}. The correlator is further constrained by the superconformal Ward identities \cite{Dolan:2004mu}:
\es{ward}{
\left[z\partial_z -  \frac12\alpha \partial_\alpha\right] \mathcal{G}_{p}(z,\bar{z};\alpha, \bar{\alpha})|_{\alpha = \frac1z} =
\left[\bar{z}\partial_{\bar{z}} -  \frac12{\bar\alpha} \partial_{\bar{\alpha}}\right] \mathcal{G}_{p}(z,\bar{z};\alpha, \bar{\alpha})|_{\bar{\alpha}=\frac{1}{\bar{z}}} &= 0\,,
}
where $z,\bar z$ and $\alpha,\bar\alpha$ are written in terms of $U,V$ and $\sigma,\tau$, respectively, as
\es{UVtozzbar}{
U=z\bar z\,,\quad V=(1-z)(1-\bar z)\,,\qquad\qquad \sigma=\alpha\bar\alpha\,,\quad \tau=(1-\alpha)(1-\bar\alpha)\,.
}
We can satisfy these constraints by expanding $ \mathcal{G}_{p}$ in superconformal blocks as
\es{SBDecomp}{
     \mathcal{G}_p(U,V;\sigma,\tau)=\sum_{\mathcal{M}\in\mathfrak{osp}(8|4)}\lambda^2_\mathcal{M}\mathfrak{G}_\mathcal{M}(U,V;\sigma,\tau)\,,
}
where ${\cal M}$ runs over all the supermultiplets appearing in the $S_2 \times S_2$ OPE, and the $\lambda^2_{\mathcal{M}}$ are the squared OPE coefficients for each such supermultiplet $\mathcal{M}$.  In Table \ref{opemult}, we list the multiplets $\mathcal{M}$ that appear in the OPE $S_2\times S_2$, the dimension $\Delta$, spin $\ell$, and $\mathfrak{so}(8)_R$ representation $[0ab0]$ of their primaries, along with the possible values of their Lorentz spins.  The superblocks are written as linear combinations of conformal blocks
 \es{GExpansion}{
  {\mathfrak G}_{\mathcal{M}}(U, V; \sigma, \tau) = \sum_{a=0}^2 \sum_{b = 0}^a Y_{ab}(\sigma, \tau)  \sum_{(\Delta',j')\in\mathcal{M}} \mathfrak{a}^\mathcal{M}_{ab \Delta' j'}(\Delta, j) G_{\Delta',j'}(U,V) \,,
 }  
where the explicit coefficients $ \mathfrak{a}^\mathcal{M}_{ab \Delta' j'}(\Delta, j) $  are given in \cite{Chester:2014fya}. The quadratic polynomials $Y_{ab}(\sigma, \tau)$ are eigenfunctions of the $\mathfrak{so}(8)$ Casimir corresponding to the various irreducible $\mathfrak{so}(8)$ representations appearing in the product ${\bf 35}_c \otimes {\bf 35}_c$, and are given by \cite{Dolan:2003hv,Nirschl:2004pa}
 \es{polyns}{
   {\bf 1} = [0000]: \qquad Y_{00}(\sigma, \tau) &= 1 \,, \\
    {\bf 28} = [0100]: \qquad Y_{10}(\sigma, \tau) &= \sigma - \tau \,, \\
   {\bf 35}_c = [0020]: \qquad  Y_{11}(\sigma, \tau) &= \sigma + \tau -\frac{1}{4} \,, \\
   {\bf 300} = [0200]: \qquad Y_{20}(\sigma, \tau) &= \sigma^2 + \tau^2 - 2\sigma\tau - \frac{1}{3}(\sigma + \tau) + \frac{1}{21} \,,\\
    {\bf 567}_c = [0120]: \qquad Y_{21}(\sigma, \tau) &= \sigma^2 - \tau^2 - \frac{2}{5}(\sigma - \tau) \,,\\
    {\bf 294}_c = [0040]: \qquad Y_{22}(\sigma, \tau) &= \sigma^2 + \tau^2 + 4\sigma\tau - \frac{2}{3}(\sigma+\tau) + \frac{1}{15} \,.
 } 
 We will sometimes find it useful to write the correlator in a basis of these polynomials as
 \es{Ybasis}{
\mathcal{G}_{p}(U,V;\sigma,\tau)=\sum^{2}_{n=0}\sum_{m=0}^{2}Y_{[0\,m\, 2n-2m\,0]}(\sigma, \tau) \mathcal{G}_{[0\,m\, 2n-2m\,0]}(U,V)\,,
}
where the coefficients $\mathcal{G}_{[0\,m\, 2n-2m\,0]}(U,V)$ can then be fixed using the superconformal Ward identity as above.

\begin{table}
\centering
\begin{tabular}{|c|c|r|c|c|}
\hline
Type    & $(\Delta,\ell)$     & $\mathfrak{so}(8)_R$ irrep  &spin $\ell$  \\
\hline
$(B,+)$ &  $(2,0)$         & ${\bf 294}_c = [0040]$& $0$  \\ 
$(B,2)$ &  $(2,0)$         & ${\bf 300} = [0200]$& $0$  \\
$(B,+)$ &  $(1,0)$         & ${\bf 35}_c = [0020]$ & $0$ \\
$(A,+)$ &  $(\ell+2,\ell)$       & ${\bf 35}_c = [0020]$ &even  \\
$(A,2)$ &  $(\ell+2,\ell)$       & ${\bf 28} = [0100]$ & odd  \\
$(A,0)$ &  $\Delta\ge \ell+1$ & ${\bf 1} = [0000]$ & even \\
$\text{Id}$ &  $(0,0)$ & ${\bf 1} = [0000]$ & even \\
\hline
\end{tabular}
\caption{The possible superconformal multiplets in the $S_2 \times S_2$ OPE.  The $\mathfrak{so}(3, 2) \oplus \mathfrak{so}(8)_R$ quantum numbers are those of the superconformal primary in each multiplet.}
\label{opemult}
\end{table}

\subsection{Tree level data}
\label{tree}

We now consider the large $c_T$ expansion of $\cG_{p}$ as well as the double trace long multiplet CFT data:
\es{Hlarge}{
\cG_{p}(U,V;\sigma,\tau)&=\cG^{(0)}_{p}+c_T^{-1}\cG^R_{p}+c_T^{-\frac53}\cG^{R^4}_{p}+\dots\,\\
\Delta_{t,\ell,I}&=t+\ell+c_T^{-1}\gamma^{R}_{t,\ell,I}+c_T^{-\frac53}\gamma^{R^4}_{t,\ell,I}+\dots\,,\\
(\lambda_{p,t,\ell,I})^2&=(\lambda^{(0)}_{p,t,\ell,I})^2+c_T^{-1}(\lambda^{R}_{p,t,\ell,I})^2+c_T^{-\frac53}(\lambda^{R^4}_{p,t,\ell,I})^2+\dots\,,\\
}
where the GFFT term is simply $\cG^{(0)}_{p}=1$. The index $I=1,\dots t-1$ denotes the double trace operators that are degenerate at $c_T\to\infty$, since they can be built from different single traces in $t-1$ different ways. A similar expansion exists for the OPE coefficients of the protected operators, although of course their scaling dimensions are fixed. Using these expansions, we can write the superblock expansion for $\cG_{p}$ in \eqref{SBDecomp} at large $c_T$ as
\es{SGexp}{
&\cG_{p}^R(U,V)={128p} \mathfrak{G}_\text{Stress}(U,V;\sigma,\tau) +\hspace{-.2in}\sum_{\mathcal{M}_{\Delta,\ell}\in\{(B,+),(B,2),(A,2)_\ell,(A,+)_\ell\}}\hspace{-.2in}\lambda^R_{22\mathcal{M}} \lambda^R_{pp\mathcal{M}}  \mathfrak{G}_\mathcal{M}(U,V;\sigma,\tau) \\
&\quad+\sum_{ t,\ell,I} \left[\lambda^{R}_{2,t,\ell,I} \lambda^{R}_{p,t,\ell,I}+\lambda^{(0)}_{2,t,\ell,I} \lambda^{(0)}_{p,t,\ell,I}\gamma_{t,\ell,I}^R(\partial_t^\text{no-log}+\frac12\log U)\right]  \mathfrak{G}_{t+\ell,\ell}(U,V;\sigma,\tau) \,.
}
Here, the first line includes the protected multiplets, and the OPE coefficient for the stress tensor multiplet was written in terms of $c_T$ using the conformal Ward identity \cite{Alday:2021ymb}. In the second line we denote the singlet long multiplet superblock by $\mathfrak{G}_{\Delta,\ell}$ and $\partial_t^\text{no-log}  \mathfrak{G}_{t+\ell,\ell}(U,V;\sigma,\tau) $ denotes that we consider the term after taking the derivative that does not include a $\log U$, which has already been written separately. The expansion of $\cG_{p}^{R^4}$ takes the same form with $R\to R^4$, except the stress tensor block does not appear.

While the superblock expansion is best expressed in position space, in the large $c_T$ expansion it is also useful to consider the Mellin transform $M_p(s,t;\sigma,\tau)$ of the connected correlator $\mathcal{G}^\text{con}_{p}(U,V;\sigma,\tau)\equiv\mathcal{G}_{p}(U,V;\sigma,\tau)-\mathcal{G}^{(0)}_{p}(U,V;\sigma,\tau)$, which is defined as \cite{Zhou:2017zaw}:
\es{mellinH}{
\mathcal{G}^\text{con}_{p}(U,V;\sigma,\tau)&=\int\frac{ds\, dt}{(4\pi i)^2} U^{\frac s2}V^{\frac t2-\frac{p}{4}-\frac12}M_p(s,t;\sigma,\tau) \\
&\qquad\qquad\times\Gamma\left[\frac p2-\frac s2\right]\Gamma\left[1-\frac s2\right]\Gamma^2\left[\frac p4+\frac12-\frac t2\right]\Gamma^2\left[\frac p4+\frac12-\frac {{u}}{2}\right],\\
}
where $u = p+2 - s - t$ and the integration contours here is defined to include all poles of the Gamma functions on one side of the contour. The Mellin amplitude is defined such that a bulk contact Witten diagram coming from a vertex with $2m$ derivatives gives rise to a polynomial in $s,t$ of degree $m$, and similarly an exchange Witten diagrams corresponds to a Mellin amplitude with poles for the twists of each exchanged operator. The Mellin amplitude must also obey the crossing relations
 \es{crossM}{
 M_p(s,t;\sigma,\tau)  =  M_p(s,u;\tau,  \sigma)   \,, \qquad   M_2(s,t;\sigma,\tau) =    \tau^2M_2(t,s;\sigma/\tau,1/\tau)\,,
 }
which follow from interchanging the first and second operators, and, for $p = 2$, the first and third. Lastly, $M_p(s,t;\sigma,\tau)$ must satisfy the Ward identities \eqref{ward}, which can be implemented in Mellin space as shown in \cite{Zhou:2017zaw}. Using all these constraints, $M_p(s, t)$ can be expanded similar to the position space expression \eqref{Hlarge} to get
\es{Mplarge}{
M_{p}(s,t)=c_T^{-1}M^R_{p}+c_T^{-\frac53}B^{R^4}(p)M^{R^4}_{p}+\dots\,,
}
where $M^{R^4}_{p}$ is a complicated degree 4 polynomial in $s,t$ given in \cite{Alday:2021ymb}, whose explicit expression we put in the attached \texttt{Mathtematica} file. Note that $M^{R^4}_2$ is the same as $M^4$ in \eqref{M2222}. The tree level supergravity amplitude $M^R_{p}$ was written in \cite{Alday:2020dtb} as an infinite sum of the supergravity multiplet and its descendents:
\es{Ms}{
M^R_p=& \sum_{m=0}^\infty\Bigg[  \frac{2^{2 m+5} p ((p-2 t+2) (p+2 (s+2) \sigma -2 (s+t-1))-2 (s+2) \tau  (p-2 (s+t-1)))}{\pi ^2 (2 m+3) (2 m-s+1) \Gamma
   \left(\frac{1}{2}-m\right) \Gamma (2 m+2) \Gamma \left(\frac{1}{2} (-2 m+p-1)\right)} \\
   &+\frac{16 p \tau  \Gamma \left(\frac{p}{2}+1\right) ((p+2 t+2) (p (2 \sigma -1)+2 (-\sigma  s+s+t-1))+2 \tau  (p-s) (p-2
   (s+t-1)))}{\pi  \Gamma \left(\frac{1}{2}-m\right)^2 \Gamma (m+1) \Gamma \left(\frac{p-1}{2}\right) (4 m+p-2 t) \Gamma
   \left(m+\frac{p+3}{2}\right)}\\
   &+\frac{2^{2 m+5} p ((p-2 t+2) (p+2 (s+2) \sigma -2 (s+t-1))-2 (s+2) \tau  (p-2 (s+t-1)))}{\pi ^2 (2 m+3) (2 m-s+1) \Gamma
   \left(\frac{1}{2}-m\right) \Gamma (2 m+2) \Gamma \left(\frac{1}{2} (-2 m+p-1)\right)}\Bigg]\,,
}
where the overall coefficient was fixed by extracting the stress tensor OPE coefficient, and so is universal for any holographic dual.\footnote{Note that this expression is degree one in $\sigma,\tau$, which is because the correlator is fixed from the graviton exchange that only contributes to the irreps $[0000]$, $[0100]$, and $[0020]$, which are degree one.} As shown in \cite{Alday:2021ymb},
for even $p$ we can perform the Mellin space integral in \eqref{mellinH} to get a position space expression that can written in terms of a finite sum of $\bar{D}_{r_1,r_2,r_3,r_4}(U,V)$ functions. These expressions were then expanded in superblocks to get the average anomalous dimensions $\langle \lambda^{(0)}_{2,t,\ell} \lambda^{(0)}_{p,t,\ell}\gamma_{t,\ell}\rangle\equiv\sum_{ I} \lambda^{(0)}_{2,t,\ell,I} \lambda^{(0)}_{p,t,\ell,I}\gamma_{t,\ell,I}$ weighted by OPE coefficients for general $t$ and $\ell$ and even $p\leq36$, as given in \cite{Alday:2021ymb}. For odd $p$, we cannot write  \eqref{mellinH} as a finite sum of $\bar{D}_{r_1,r_2,r_3,r_4}(U,V)$ functions. Instead, we do the Mellin transform in a small $U$ and $1-V$ expansion, then resum the $V$ expansion using the ansatz
 \es{p2A22R4}{
\mathcal{G}^{R}_{[0040]}(U,V)&=\
\sum_{t=3,5,\dots}{U^{\frac t2}} \Big[P^t_1(V)+P^t_2(V)E(1-V)+P^t_3(V)K(1-V)\Big]\,,
 }
 where we used the basis \eqref{Ybasis}, and similar expressions exist for the other channels. Here, $P^t_i(V)$ are polynomials in $V$ divided by monomials in $(1-V)$, and $K(1-V)$ and $E(1-V)$ are elliptic integrals of the first and second kind, respectively, which occur in sums of odd twist data as discussed in \cite{Aharony:2018npf}.  We can then easily expand in superblocks to get the average odd twist OPE coefficients $\langle (\lambda^{R}_{p,t,\ell})^2\rangle$ for odd $p$ and $t$ and any $\ell$. Although we could not guess any simple closed form for this data, we computed enough to get the 1-loop expression in the next subsection. We give some sample values in the attached \texttt{Mathematica} file.

For the $R^4$ amplitude, we need to fix the overall coefficient $B^{R^4}(p)$, which depends on the specific holographic dual. For $AdS_4\times S^7/\mathbb{Z}_k$, it was shown in \cite{Alday:2021ymb} that the flat space limit fixes 
 \es{BR4}{
B^{R^4}(p)=\frac{32\cdot{2^{\frac13}} (p-1) (p+1) (p+3) (p+5)}{3\cdot 3^{\frac23} \pi ^{8/3} k^{2/3} \Gamma \left(\frac{p}{2}\right)}\,.
}
Since $M_p^{R^4}$ is just a polynomial Mellin amplitude for every $p$, it is straightforward to take the Mellin transform to get \cite{Alday:2021ymb} 
 \es{p2A22R42}{
\mathcal{G}^{R^4}_{[0040]}(U,V)&=\
 \frac{64 \cdot{2^{\frac13}} \left(p^2-1\right) U^{p/2}} { 3^{8/3} \pi ^{8/3} k^{2/3} \Gamma \left(\frac{p}{2}\right)}
  \Big[6 (p-2) p \bar{D}_{\frac{p}{2},\frac{p}{2},1,1}(U,V)+4 (p-3) (p-2) (p+2)
   \bar{D}_{\frac{p}{2},\frac{p}{2},2,2}(U,V)\\
   &+192 \bar{D}_{\frac{p}{2},\frac{p}{2},3,3}(U,V)-312
   \bar{D}_{\frac{p}{2},\frac{p}{2},4,4}(U,V)+ 
  p (p-1) ((p-1) p-26)
   \bar{D}_{\frac{p}{2},\frac{p}{2},3,3}(U,V)\\
   &+4 p(p (p+2)-29) \bar{D}_{\frac{p}{2},\frac{p}{2},4,4}(U,V)+4p (p+8)
   \bar{D}_{\frac{p}{2},\frac{p}{2},5,5}(U,V)+60 \bar{D}_{\frac{p}{2},\frac{p}{2},5,5}(U,V)
   \Big]\,,
 }
 and similarly for the other channels. We can then expand in superblocks to get the tree level data for even and odd $p$:
\es{averageR4}{
\text{even $p$}:&\quad \langle \lambda^{(0)}_{2,t,\ell} \lambda^{(0)}_{p,t,\ell}\gamma^{R^4}_{t,\ell}\rangle=
\delta_{\ell,0}(-1)^{\frac p2} \frac{ (t+1) 2^{p+t+\frac{25}{3}} \Gamma \left(\frac{t}{2}+2\right) \Gamma \left(\frac{t}{2}+4\right) \Gamma \left(\frac{1}{2} (p+t+5)\right)}{ 3^{5/3} \pi ^{8/3}
   k^{2/3} \Gamma (p-1) \Gamma \left(t+\frac{5}{2}\right) \Gamma \left(\frac{1}{2} (-p+t+2)\right)}
\,,\\
\text{odd $p$}:&\qquad\quad\; \langle (\lambda^{R^4}_{p,t,\ell})^2\rangle=\delta_{\ell,0}(-1)^{\frac{p+3}{2}} \frac{\pi}{2} \frac{ (t+1) 2^{p+t+\frac{25}{3}} \Gamma \left(\frac{t}{2}+2\right) \Gamma \left(\frac{t}{2}+4\right) \Gamma \left(\frac{1}{2} (p+t+5)\right)}{ 3^{5/3} \pi ^{8/3}
   k^{2/3} \Gamma (p-1) \Gamma \left(t+\frac{5}{2}\right) \Gamma \left(\frac{1}{2} (-p+t+2)\right)}  \,,
}
where the even $p$ anomalous dimensions were already computed in \cite{Alday:2021ymb}, and note that the average OPE coefficients are very similar to the average anomalous dimensions. The fact that we only have spin zero is expected from the classic discussion in \cite{Heemskerk:2009pn}

\subsection{1-loop data}
\label{1fromtree}

We now use this tree and GFFT data to compute the double discontinuity (DD) of the 1-loop correlator, which is sufficient to extract CFT data for sufficiently large spin. We start by expanding the correlator $\cG$ for $\langle2222\rangle$ to 1-loop order at large $c_T$ using the superblock expansion. For $R|R$ at order $c_T^{-2}$, this takes the form
\es{RR}{
\cG^{R|R}=&\sum_{t=2,4,\dots}\sum_{\ell\in\text{Even}}\Big[\frac18\langle(\lambda^{(0)}_{t,\ell})^2(\gamma^R_{t,\ell})^2\rangle(\log^2U+4\log U\partial_t^\text{no-log}+4(\partial_t^\text{no-log})^2) \\
&+\frac12\langle(\lambda^{R})^2_{t,\ell}\gamma^{R}_{t,\ell}\rangle(\log U+2\partial_t^\text{no-log})\\
 &+\frac12\langle(\lambda^{(0)}_{t,\ell})^2\gamma^{{R}|{R}}_{t,\ell}\rangle(\log U+2\partial_t^\text{no-log})+\langle(\lambda^{{R}|{R}}_{t,\ell})^2\rangle\Big] \mathfrak{G}_{t+\ell,\ell}(U,V;\sigma,\tau)\\
 &+\sum_{\mathcal{M}_{\Delta,\ell}\in\{(B,+),(B,2),(A,2)_\ell,(A,+)_\ell\}}(\lambda^{R|R}_{22\mathcal{M}} )^2   \mathfrak{G}_\mathcal{M}(U,V;\sigma,\tau)\,,
}
where $\partial_t^\text{no-log} \mathfrak{G}_{t+\ell,\ell}(U,V;\sigma,\tau)$ was defined in \eqref{SGexp}. The first three lines describe the double trace singlet long multiplets $(A,0)^{[0000]}_{t+\ell,\ell}$, where $\langle\rangle$ denotes the average over the $ (t-1)$-fold degenerate operators. The fourth line includes all the protected multiplets in $\langle 2222\rangle$ except the stress tensor multiplet, which is $1/c_T$ exact. The expression for $\cG^{R^4|R^4}$ at order $c_T^{-\frac{10}{3}}$ is identical except we replace $R\to R^4$ and the sum for the long multiplets is now restricted to $\ell=0$, while for $\cG^{R|R^4}$ at order $c_T^{-\frac83}$ we furthermore replace the $\frac18$ in the first line by $\frac14$, since the vertices are different.

The DD comes from singularities as $V\to0$. To find these terms we perform $1\leftrightarrow3$ crossing 
\es{crossing}{
\cG(U,V;\sigma,\tau)=\frac{U}{V}\tau^2 \cG(V,U;\sigma/\tau,1/\tau)\,,
}
and find that only $\log^2 U$ or odd twist terms in the $s$-channel can contribute to the DD after crossing. For 1-loop vertices $A,B=R,R^4$, the $\log^2 U$ terms multiply the average $\langle(\lambda^{(0)}_{t,\ell})^2\gamma^A_{t,\ell}\gamma^{B}_{t,\ell}\rangle$, which can be computed by unmixing the GFFT $\langle ppqq\rangle$ and tree level $\langle22pp\rangle$ data for each KK mode in the theory:
\cite{Alday:2017xua,Aprile:2017bgs,Aprile:2017xsp,Alday:2018pdi}
\es{appA}{
\langle(\lambda^{(0)}_{t,\ell})^2\gamma^A_{t,\ell}\gamma^{B}_{t,\ell}\rangle=\sum_{p=2,4,\dots}^{t}\frac{\langle\lambda^{(0)}_{2,t,\ell}\lambda^{(0)}_{p,t,\ell}\gamma^A_{t,\ell}\rangle  \langle\lambda^{(0)}_{2,t,\ell}\lambda^{(0)}_{p,t,\ell}\gamma^B_{t,\ell}\rangle}{ {\langle(\lambda^{(0)}_{p,t,\ell})^2\rangle} }\,,
}
where we summed over each $p$ for which a given twist $t$ long multiplet appears. Note that this sum only runs over even $p$, and so is the same for the $k=1,2$ theories. This contribution to the DD is similar to the 4d \cite{Alday:2017xua} and 6d \cite{Alday:2020tgi} cases, and is in fact the only contribution for the $k=2$ theory considered in \cite{Alday:2021ymb}. For the $k=1$ case we consider now, we must also consider odd twist terms in the $s$-channel, which under crossing will have factors of $\sqrt{V}$ that contribute to the DD. These terms multiply the average $\langle(\lambda^{{A}|{B}}_{t,\ell})^2\rangle$ for odd $t$, which can be similarly computed by unmixing GFFT and tree data as
\es{appA2}{
\langle(\lambda^{{A}|{B}}_{t,\ell})^2\rangle=\sum_{p=3,5,\dots}^{t} \frac{ \langle (\lambda^{A}_{p,t,\ell})^2\rangle \langle (\lambda^{B}_{p,t,\ell})^2\rangle   }{ {\langle(\lambda^{(0)}_{p,t,\ell})^2\rangle} }\,,
}
where the 1-loop odd OPE coefficients in $\langle2222\rangle$ are related to tree level OPE coefficients in $\langle22pp\rangle$, because the latter correlator does not contain any GFFT term for these operators. Note that these odd operators do not exist in 6d and 4d, and were projected out in the $k=2$ theory considered in \cite{Alday:2021ymb}. 

We can now use the average tree level data from the previous section, as well as ${\langle(\lambda^{(0)}_{p,t,\ell})^2\rangle}$ as given in \cite{Alday:2021ymb} and give in the \texttt{Mathematica} file, to compute the DD for $R|R$, $R|R^4$, and $R^4|R^4$. For the average anomalous dimension in all of these theories, we can perform the $p,t,\ell$ sums by expanding at small $U$ in each $R$-symmetry channel to get:
\es{slices2}{
\frac18\sum_{t=2,4,\dots}& \sum_{\ell\in\text{Even}}\sum_{p=2,4,\dots}^{t}  \frac{ \langle (\lambda^{A}_{p,t,\ell})^2\rangle \langle (\lambda^{B}_{p,t,\ell})^2\rangle   }{ {\langle(\lambda^{(0)}_{p,t,\ell})^2\rangle} }\mathfrak{G}_{t+\ell,\ell}(U,V;\sigma,\tau)=\\
& Y_{[0000]}(\sigma,\tau) \sum_{t=2,4,\dots}{U^{\frac t2}} \Big[Q^t_1(V)+Q^t_2(V)\log V+Q^t_3(V)\log^2V+Q^t_4(V) \text{Li}_2(1-V)\Big]+\dots\,,
}
where the dots denote the other channels that will start at higher powers of $U$, and $Q_i(V)$ are polynomials of $V$ divided by monomials in $(1-V)$. For 1-loop terms with an $R^4$ vertex, the $Q_3(V)$ and $Q_4(V)$ terms vanish. These $U$ slices for $AdS_4\times S^7$ are identical to those of $AdS_4\times S^7/\mathbb{Z}_2$ as given in \cite{Alday:2021ymb}, except that for the $R^4$ terms we should be careful about the factor of $k^{-2/3}$ in \eqref{averageR4}. For the average OPE coefficient contribution to $AdS_4\times S^7$, we can perform the $p,t,\ell$ sums by expanding at small $U$ in each $R$-symmetry channel to get:
\es{slices}{
\frac18\sum_{t=3,5,\dots}& \sum_{\ell\in\text{Even}}\sum_{p=3,5,\dots}^{t}  \frac{ \langle (\lambda^{A}_{p,t,\ell})^2\rangle \langle (\lambda^{B}_{p,t,\ell})^2\rangle   }{ {\langle(\lambda^{(0)}_{p,t,\ell})^2\rangle} }\mathfrak{G}_{t+\ell,\ell}(U,V;\sigma,\tau)=\\
& Y_{[0000]}(\sigma,\tau) \sum_{t=3,5,\dots}{U^{\frac t2}} \Big[R_1(V)+R_2(V)E(1-V)+R_3(V)K(1-V)\Big]+\dots\,,
}
where the dots denote the other channels that will start at higher powers of $U$, and note that the ansatz for the $V$ dependence is the same as \eqref{p2A22R4}. For 1-loop terms with an $R^4$ vertex, the $R_1(V)$ terms vanish.

We can now perform $1\leftrightarrow3$ crossing on these expressions as in \eqref{crossing} and then apply the Lorentzian inversion formula to the resulting DDs to extract CFT data. These 1-loop inversion formulae were already computed in \cite{Alday:2021ymb}, the only difference now is that the DD of the even and odd twist contributions will give different factors. For instance, we can compute the 1-loop $\lambda^2_{(A,+),\ell} $ as
\es{ApInversion}{
\lambda^2_{(A,+),\ell} =\frac{12 (2 \ell+5) \Gamma (\ell+3)^4}{\Gamma \left(\ell+\frac{5}{2}\right)^2
   \Gamma \left(\ell+\frac{7}{2}\right)^2} \int_0^1 \frac{d \bar z}{\bar z}    g_{\ell+4,\ell+2}(\bar z) \text{dDisc}[ {\cal G}^{[0040]}(z\bar z, 1-\bar z)\vert_z ] \,,
}
where we set $U=z\bar z \,, V=(1-z)(1-\bar z)$, and the DD of the even and odd twists are computed as
\es{DD}{
{\rm dDisc}\,[ f(z,\bar z)\sqrt{1-\bar z} ] = 2 \sqrt{1-\bar z} f(z, \bar z)\,,\qquad {\rm dDisc}\,[ f(z,\bar z)\log^2{1-\bar z} ] = 4\pi^2 f(z,\bar z)\,,
}
where we assume here that $f(z,\bar z)$ is analytic as $\bar z\to1$. Similar formulae are given for the other CFT data in \cite{Alday:2021ymb}. For the lowest-twist $R|R$ anomalous dimensions, the inversion formula must include the normalization correction $(A_{2\ell}-7/12)(\gamma^R_{2,\ell})^2$, where $A_n=\sum_{k=1}^n(-1)^{k+1}/k$ and $A_0=0$. This accounts for the dimension dependence of the conformal-block normalization and the superblock coefficient, and vanishes for $\ell=2$. The resummed 1-loop DD's in the relevant irreps are given in the attached \texttt{Mathematica} file. We can then apply the inversion formulae to these DDs to get the $R|R$ CFT data:
\es{RRfinal}{
 (\lambda^{R|R}_{(A,+)_0})^2&=513.49235\,,\\
(\lambda^{R|R}_{(A,+)_2})^2&=111.509682\,,\\
(\lambda^{R|R}_{(A,+)_4})^2&=63.56569\,,\\
(\lambda^{R|R}_{(A,2)_1})^2&=4170.0039549\ldots\,,\\
(\lambda^{R|R}_{(A,2)_3})^2&=836.113018\,,\\
(\lambda^{R|R}_{(A,2)_5})^2&=468.97184671\ldots\,,\\
\gamma_{2,2}^{R|R}&=\frac{1645242368}{1125 \pi ^4}-\frac{1775517856}{3315 \pi ^2}\,,\\
\gamma_{2,4}^{R|R}&=\frac{80952501344}{25725 \pi ^4}-\frac{2328151696}{6783 \pi ^2}\,,\\
}
where the Lorentzian inversion formula converges for all the data listed here (and higher spins and twists), which is only missing $\gamma_{2,0}^{R|R}$, $(\lambda^{R|R}_{(B,+)})^2$, and $(\lambda^{R|R}_{(B,2)})^2$. For the $R|R^4$ term, we can similarly compute
\es{RR4final}{
(\lambda^{R|R^4}_{(A,+)_4})^2&=\frac{44583908016128 \cdot 2^{1/3}}{4357815\cdot 3^{\frac23} \pi ^{14/3}}+\frac{12490452091535360 \cdot 2^{1/3}}{6298655363\cdot 3^{\frac23} \pi ^{8/3}}\,,\\
(\lambda^{R|R^4}_{(A,2)_5})^2&=\frac{509628037520687104 \cdot 2^{1/3}}{3277699425\cdot 3^{\frac23} \pi ^{14/3}}+\frac{2251799813685248 \cdot 2^{1/3}}{72177105\cdot 3^{\frac23} \pi
   ^{8/3}}\,,\\
\gamma_{2,6}^{R|R^4}&=-\frac{1025280925696 \cdot 2^{1/3}}{693\cdot 3^{\frac23} \pi ^{14/3}}-\frac{295081030451200 \cdot6^{1/3}}{2956811 \pi ^{8/3}}\,,\\
}
where here the Lorentzian inversion formula does not converge for lower spins.

\subsection{Mellin amplitudes and the flat space limit}
\label{flatSec}

We can compute the Mellin amplitudes from the resummed DD's following \cite{Alday:2021ymb}. As discussed, the even twist contributions to $AdS_4\times S^7$ are identical to those of $AdS_4\times S^7/\mathbb{Z}_2$, up to overall factors of $2$ for the $R^4$ vertices, so we will only need to compute the odd twist contributions then add them to the results of \cite{Alday:2021ymb}.

In the previous subsection, we computed the coefficient of $\sqrt{U}$ in the $s$-channel, which gave the DD in the $t$-channel as an expansion in small $U$. We can use the definition of the Mellin amplitude in \eqref{mellinH} to convert these terms to poles in $s$ with a residue in $t$, and then use crossing symmetry \eqref{crossM} to fix the other parts of the Mellin amplitude that are analytic in $s$. For $R|R^4$ and $R^4|R^4$ both expressions take the form
\es{RR4c}{
M_\text{odd}^{A|R^4}(s,t;\sigma,\tau) &= \sum_{m=3/2}^\infty\Bigg[ \frac{\hat{d}(m,s,t;\sigma,\tau)\Gamma(m)}{(s-2m)\Gamma(m+\frac12)} +\text{crossed}    \Bigg]+\sum_{i=1}^{n} \hat k_i{\bf P}^{i}(s,t;\sigma,\tau)\,,\\
}
where $\hat {d}(m,s,t;\sigma,\tau)$ are quadratic in $\sigma,\tau$ and polynomials in $m,s,t$, while ${\bf P}^{i}(s,t;\sigma,\tau)$ parameterize all $n=50$ and $n=84$ crossing symmetric polynomials in $s,t$ of degree 8 and 11 for $R|R^4$ and $R^4|R^4$, respectively. The coefficients $\hat k_i$ can be fixed by the superconformal Ward identity in terms of a smaller subset of physical contact term ambiguities. The $\hat {d}(m,s,t;\sigma,\tau)$ are actually identical to the coefficients in the even twist contribution to $M^{R|R^4}$ and  $M^{R^4|R^4}$ in \cite{Alday:2021ymb}, except $M^{R|R^4}$ contains an extra term in the sum with a different $\Gamma$ factor, and the $m$ sum runs over integers. Note that one can swap $s$ for $2m$ in these expressions to get the same residues at the poles, which only changes $\hat k_i$. The only rule in performing this swap is that the degree of $M_\text{odd}^{R|R^4}(s,t) $ and $M_\text{odd}^{R^4|R^4}(s,t) $ at large $s,t$ does not exceed $8.5$ and $11.5$, respectively. In practice we can simply set $s=2m$ and similarly for the crossed terms, which allows us to resum to get
\es{RR4c2}{
M_\text{odd}^{A|R^4} &=\Bigg[  \Bigg(\frac{\sqrt{\pi }}{s-1}+\frac{\pi  \Gamma \left(\frac{1}{2}-\frac{s}{2}\right)}{2 \Gamma
   \left(1-\frac{s}{2}\right)} \Bigg) {\bf p}(s,t;\sigma,\tau)+\text{crossed}\Bigg]+\sum_{i=1}^{n} \hat k_i{\bf P}^{i}(s,t;\sigma,\tau)\,,\\
}
where ${\bf p}(s,t;\sigma,\tau)$ and ${\bf P}^{i}(s,t;\sigma,\tau)$ are polynomials in $s,t,\sigma,\tau$, and the $\hat k_i$ can be fixed using the Mellin space Ward identity. In the attached \texttt{Mathematica} file we give the final complete expression as a function of $s,t,\sigma,\tau$ after fixing all these coefficients.

For $R|R$, the $R_1(V)$ term in the slices \eqref{slices} is now non-vanishing, which gives a Mellin amplitude with both double and single poles in $s,t$:
\es{MellinRR}{
M_\text{odd}^{R|R}(s,t;\sigma,\tau) &= \Bigg[ \sum_{m,n=3/2}^\infty \frac{c(m,n,s,t;\sigma,\tau)}{(s-2m)(t-2n)}\frac{\Gamma(m)\Gamma(n)\Gamma(m+n-\frac{11}{2})}{\Gamma(m+n-1)\Gamma(n-\frac12)\Gamma(m-\frac12)} \\
&  + \sum_{m=3/2} \frac{1}{s-2m}\frac{d(m,s,t;\sigma,\tau)\Gamma(m)}{\Gamma(m+\frac12)(m-4)(m-3)(m-2)(m-1)}\\
& +\text{crossed}\Bigg]+\hat{\bf P}(s,t;\sigma,\tau)+\sum_{i=1}^{24} k_i{\bf P}^{(5),i}(s,t;\sigma,\tau)\,.
}
Here, ${c}(m,n,s,t;\sigma,\tau)$ and $ {d}(m,s,t;\sigma,\tau)$ are quadratic in $\sigma,\tau$ and polynomials in $m,s,t$, while ${\bf P}^{(5),i}(s,t;\sigma,\tau)$ are all crossing symmetric degree 5 polynomial in $s,t$, which in principle should be fixed by the superconformal Ward identity in terms of just one of the 24 $k_i$. Again we find that the coefficient ${c}(m,n,s,t;\sigma,\tau)$ is identical to the coefficients in the even twist contribution in \cite{Alday:2021ymb}, except the sums in that case run over integer $m,n$, and the $ {d}(m,s,t;\sigma,\tau)$ coefficients slightly differ from the even twist case. For the double pole residues we can swap $s$ for $2m$ and $t$ for $2n$ to get the same residue at the poles, but which will change the single pole residues and the $k_i$. When swapping we must be careful that the resulting sums are all finite, and that the large $s,t$ growth does not exceed $5.5$. In fact, for all choices of swaps the large $s,t$ degree exceeds $5.5$, which is why we must also include the polynomial $\hat{\bf P}(s,t;\sigma,\tau)$ that generically will have degree greater than $5.5$, and is fixed to cancel the corresponding large $s,t$ terms from the single and double sum terms. 

We can take the flat space limit of these Mellin amplitudes and compare to the 11d S-matrix in \eqref{A}. The flat space limit formula \cite{Penedones:2010ue,Chester:2018aca} relates the $\langle2222\rangle$ Mellin amplitude $M^a(s,t)$ of large $s,t$ degree $a$ to the 11d amplitude defined in \eqref{A} as
 \es{flat}{
c_T^{\frac{2(1-a)}{9}}\frac{\pi ^{5/2} 2^{-a-5}}{\Gamma \left(\frac{1}{2} (2 a+1)\right)}\lim_{s,t\to\infty} \frac{s t (s+t)M^a(s,t)}{(t (-\sigma  s+s+t)+s \tau  (s+t))^2}= \ell_{11}^{2a-2}\frac{{\mathcal{A}_{2a+7}}}{\cA_R}\,,
 }
 where $\mathcal{A}_{2a+7}$ is a term in the amplitude with length dimension $(2a+7)$, and $\ell_{11}$ is the 11d Planck length. Since the only dependence on $k$ comes from the dictionary \eqref{cPlanck}, we expect that the $k=1,2$ theories should differ by just a factor of $2$ for $R|R$, with an extra factor of $2^{2/3}$ and $2^{4/3}$ for $R|R^4$ and $R^4|R^4$. At large $s,t$ we find that the odd twist contributions to the amplitudes that we computed here become identical to the even twist contributions in \cite{Alday:2021ymb}:
\es{flatRR4}{
\lim_{s,t\to\infty}M_\text{odd}^{A|B} (s,t;\sigma,\tau)&=\lim_{s,t\to\infty}M_\text{even}^{A|B} (s,t;\sigma,\tau) \,,\\
}
for $A,B=R,R^4$. This can be seen from the fact that the coefficients of the leading terms in the sums in \eqref{RR4c} and \eqref{MellinRR} are identical to the even twist case,\footnote{For $R|R$, it is easiest to compare the formulation of \eqref{MellinRR} with the summand multiplied by $mn/(s,t)$, which has an easier flat space limit without the need for the polynomial term $\hat{\bf P}(s,t;\sigma,\tau)$. See \cite{Alday:2021ymb} for analogous details in the $k=2$ case.} and at large $s,t$ it does not matter if one sums over integer or half integer $m,n$. Since we already checked that the $k=2$ theory precisely matched the 11d amplitude in \cite{Alday:2021ymb}, this implies that the $k=1$ theory also matches.

\section{Constraints from supersymmetric localization}
\label{protOPE}

The calculations so far have only used the general constraints of the analytic bootstrap, and not the detailed Lagrangian of the interacting sector of $U(N)_1\times U(N)_{-1}$ ABJM theory, which is dual to M-theory on $AdS_4\times S^7$. We will now use results from supersymmetric localization, which uses the Lagrangian, to further constrain correlation functions. We start by using the recent results of \cite{Gaiotto:2020vqj} to compute OPE coefficients to all orders in $1/N$ for short operators that appear in both $\langle 22 pp\rangle$ as well as $\langle 3333\rangle$. We then use these results for $\langle2222\rangle$ to fix the contact term in the 1-loop correlator $\mathcal{G}^{R|R}$, which we find to nontrivially vanish. For $\mathcal{G}^{R|R^4}$, we combine this constraint with the other localization constraint of \cite{Binder:2018yvd} to give evidence that the conjectured analytic continuation of the Lorentzian inversion for the $k=2$ theory in \cite{Alday:2021ymb} also applies to the $k=1$ theory, which implies that contact terms also vanish for this correlator.

\subsection{Short OPE coefficients to all orders in $1/N$}
\label{flat11d}

In \cite{Kapustin:2009kz}, supersymmetric localization was used to compute the mass deformed sphere free energy $F(m_i)$ for $U(N)_k\times U(N)_{-k}$ ABJM theory for all $k$ and for all three masses $m_i$ in terms of an $N^2$ dimensional integral. If one of these masses is zero, then the quantity can also be computed to all orders in $1/N$ using the Fermi gas method \cite{Marino:2011eh,Nosaka:2015iiw}.

It was shown in \cite{Agmon:2017xes} how $\partial_m^4 F\big\vert_{m=0}$ could be used to compute the short OPE coefficients $\lambda^2_{(B,+)^{0020}_{1,0}}$, $\lambda^2_{(B,+)^{0040}_{2,0}}$, and $\lambda^2_{(B,2)_{2,0}^{0200}}$ that appear in the $S_2\times S_2$ OPE for ABJM theory with any $k$. For instance, in terms of $c_T=256/\lambda^2_{(B,+)^{0020}}$, we have the all orders in $1/N$ expression for the interacting sector of $k=1$ ABJM:
\es{cTABJM}{
c_T&=-15-\frac{112}{3\pi^2}-\frac{8(9+8N)\text{Ai}'\left[\left(N-3/8\right)({\pi^2}/{2})^{1/3}\right]}{3(\pi^2/2)^{2/3} \text{Ai}\left[\left(N-{3}/{8}\right)({\pi^2}/{2})^{1/3}\right] }+O(e^{-N})\,,\\
}
where we were careful to subtract the free contribution $c_T^\text{free}=16$ from the full $U(N)_k\times U(N)_{-k}$ result. The expressions for $\lambda^2_{(B,+)^{0040}_{2,0}}$ and $\lambda^2_{(B,2)^{0200}_{2,0}}$ take a similar but more complicated form, and are reviewed in the attached \texttt{Mathematica} file. They satisfy the linear relation
\es{crossConstraints}{
\frac{1024}{c_T}- 5 { \lambda}^2_{(B,+)} +{ \lambda}^2_{(B,2)} + 16 = 0\,,
}
which was derived in \cite{Chester:2014mea} from crossing symmetry in the 1d protected topological sector.

The methods of  \cite{Agmon:2017xes} are not sufficient to compute OPE coefficients of $(B,+)$ and $(B,2)$ multiplets that appear in more general correlators. For $k=1$, however, \cite{Gaiotto:2020vqj} showed how the Fermi gas method could be applied to the 1d Lagrangian in \cite{Dedushenko:2016jxl} to compute any such short OPE coefficients to all orders in $1/N$. These OPE coefficients can be efficiently computed for the full $k=1$ theory using the \texttt{Mathematica} code attached to \cite{Gaiotto:2020vqj}, where one must carefully subtract the free theory contribution. For instance, for $\lambda^2_{(B,+)^{0040}_{2,0}}$ and $\lambda^2_{(B,2)_{2,0}^{0200}}$ in $\langle22pp\rangle$, we find the $1/c_T$ expansions
\es{22ppc}{
\lambda^2_{(B,+)^{0040}_{2,0}}&=\frac{16}{3}\delta_{p,2}+
\frac{256 (p-1) \left(\pi ^2 p^2-2 p^2 \psi
   ^{(1)}\left(\frac{p}{2}\right)+8\right)}{3 \pi ^2
 pc_T}\\
 &+\frac{4096\cdot{2}^{\frac13} p
   \left(p^2-1\right)}{3\cdot 3^{2/3} \pi ^{8/3}c_T^{5/3}}+\frac{f(p)}{c_T^2}-\frac{4096\cdot 2^{2/3}
   (p-1) p (p+1)
   (23 p (p+2)-104)}{45\cdot{3}^{\frac13} \pi
   ^{10/3 }  c_T^{7/3}}+\dots\,,\\
   \lambda^2_{(B,2)^{0200}_{2,0}}&=\frac{32}{3}\delta_{p,2}+\frac{256 \left(-\pi ^2 (p+2) p^2+2 (p+2) p^2 \psi
   ^{(1)}\left(\frac{p}{2}\right)+4 (p-1) (3
   p+4)\right)}{3 \pi ^2 pc_T}\\
 &+\frac{8192 \cdot{2}^{\frac13} \left(p^3+3 p^2-p-3\right)}{3\cdot
   3^{\frac23} \pi ^{8/3}c_T^{5/3}}+\frac{g(p)}{c_T^2}-\frac{32768\cdot 2^{\frac23}
    (p-1) (p+1)
   (p+3) (p (p+2)-4)}{9 \cdot{3}^{\frac13} \pi ^{10/3}c_T^{7/3}}+\dots\,,\\
}
where $ \psi^{(1)}\left(x\right)$ denotes the degree 1 Polygamma function, the $1/c_T$ contribution was given already in \cite{Behan:2021pzk}, and the $1/c_T^{5/3}$ term can be matched to \eqref{p2A22R42} for $k=1$, which was fixed from the flat space limit. For the $1/c_T^2$ term we could not find a simple pattern, but the $p=2$ values are
\es{1loopBs}{
f(2)=g(2)/5=-\frac{1024(400-353\pi^2+45\pi^4)}{45\pi^4}\,.
}
Note that for $p=2$ we have the linear relation \eqref{crossConstraints}, but for $p>2$ these OPE coefficients are independent. 

We also computed all the other short OPE coefficients that appear in four point functions of $S_2$ and $S_3$, which we give in the attached \texttt{Mathematica} notebook. In Table \ref{compare} we compare these all order in $1/N$ expressions to the exact expressions for the $N=3$ theory \cite{Agmon:2019imm}, which is the interacting theory with the smallest known $c_T$ that appears in the $S_3\times S_3$ OPE. We find that the all orders in $1/N$ expressions are extremely accurate even for this lowest value.

\begin{table}[htpbp]
\begin{center}
\begin{tabular}{c|c|c}
 & Exact & Large $N$   \\
 \hline
$\lambda^2_{22(B,+)_{2,0}^{[0040]}} $& 8.6761 &  8.6760  \\
 \hline
$\lambda^2_{22(B,2)_{2,0}^{[0200]}} $& 5.5934 &   5.5930 \\
 \hline
$\lambda^2_{33(B,+)_{3,0}^{[0060]}} $& 24.133 &  24.131  \\
 \hline
$\lambda^2_{33(B,2)_{3,0}^{[0220]}} $& 21.11 & 21.10   \\
 \hline
$\lambda^2_{33(B,+)_{2,0}^{[0040]}} $&12.387  &  12.385  \\
 \hline
$\lambda^2_{33(B,2)_{2,0}^{[0200]}} $& 31.945 & 31.942   \\
 \hline
$\lambda^2_{23(B,+)_{\frac52,0}^{[0050]}} $& 9.1361 & 9.1364   \\
 \hline
$\lambda^2_{23(B,2)_{\frac52,0}^{[0130]}} $&  16.8635& 16.8633   \\
 \hline
$\lambda^2_{23(B,2)_{\frac52,0}^{[0210]}} $&0 &  0.002 \\
 \hline
$\lambda_{22(B,+)_{2,0}^{[0040]}} \lambda_{33(B,+)_{2,0}^{[0040]}} $& 10.3671 &  10.3674  \\
 \hline
$\lambda_{22(B,2)_{2,0}^{[0200]}} \lambda_{33(B,2)_{2,0}^{[0200]}} $&$ -13.367$ & $-13.366$   \\
   \hline
\end{tabular}
\end{center}
\caption{Comparison of nontrivial pairs of OPE coefficients $\lambda_{pp \mathcal{M}}\lambda_{qq \mathcal{M}}$ of multiplets $\mathcal{M}$ that appear in the $\langle ppqq\rangle$ correlators for $p,q=2,3$ between the exact values computed for the interacting $N=3$ ABJM theory in \cite{Agmon:2019imm}, and the all order in $1/N$ expressions given here. We show enough digits for each quantity to show the first digit of discrepancy.
} \label{compare}
\end{table}

\subsection{Fixing 1-loop contact terms}
\label{contacts}

We can now use $\lambda^2_{(B,+)^{0040}_{2,0}}$ and $\lambda^2_{(B,2)_{2,0}^{0200}}$ to fix the contact term ambiguities in the 1-loop Mellin amplitudes, where we have one constraint because these OPE coefficients are linearly dependent for $p=2$. As was discussed for the $k=2$ theory in \cite{Alday:2021ymb}, the Lorentzian inversion formula can be analytically continued to even the low spins where naively it would not be expected to converge, and we can identify the results for this case to correspond to the 1-loop contribution with no contact term. For $R|R$, there is a single contact term we would like to fix, which in Mellin space is given by the unique degree four polynomial in $s,t,u$ that satisfies the superconformal Ward identity, whose explicit form we give in the attached \texttt{Mathematica} file. When we extract $\lambda^2_{(B,+)^{0040}_{2,0}}$ and $\lambda^2_{(B,2)_{2,0}^{0200}}$ using this inversion formula for the $R|R$ correction to the interacting $k=1$ theory, we find exactly the localization result in \eqref{22ppc} for $p=2$, which shows that the contact term vanishes. Note that we do not need the explicit Mellin amplitude to make this comparison, since only the DD is needed for the inversion formula. We can then also extract the spin zero anomalous dimension to get
\es{spin0anom}{
\gamma_{2,0}^{R|R}=\frac{39639040}{9 \pi ^4}-\frac{201244960}{429 \pi ^2}\,.
}

For $R|R^4$, there are now four contact terms we would like to fix, which are given by the four degree eight polynomials in $s,t,u$ that satisfy the superconformal Ward identity, whose explicit form we give in the attached \texttt{Mathematica} file. In addition to the single localization constraint from $\lambda^2_{(B,+)^{0040}_{2,0}}$ and $\lambda^2_{(B,2)_{2,0}^{0200}}$, there is another localization constraint that comes from the mixed mass derivative \cite{Binder:2019mpb}:
 \es{DerSimpTwoMassesFinal}{
&\frac{\partial \log Z}{\partial m_+^2 \partial m_-^2} \Big\vert_{m_\pm=0}= \frac{\pi^2 c_T^2}{2^{11}}I_{+-}[\cS^i]\,,\\
&\qquad\quad\;\;\, I_{+-}[{\cal S}^i]  \equiv \int \frac{ds\ dt}{(4\pi i)^2} \frac{2\sqrt{\pi}}{(2-t)(s+t-2)}\mathfrak{M}_1(s,t) \\
&\qquad\qquad\quad\times \Gamma \left[1-\frac{s}{2}\right] \Gamma \left[\frac{s+1}{2}\right] \Gamma \left[1-\frac{t}{2}\right] \Gamma \left[\frac{t-1}{2}\right] \Gamma \left[\frac {s+t-2}{2}\right] \Gamma \left[\frac{3-s-t}2\right]\,,
}
where $\mathfrak{M}_1(s,t)$ is the first element of the Mellin amplitude basis
\es{Mbasis}{
M(s,t;\sigma,\tau) = \mathfrak{M}_1+\sigma^2 \mathfrak{M}_2+\tau^2 \mathfrak{M}_3+\sigma\tau \mathfrak{M}_4+\tau  \mathfrak{M}_5+\sigma  \mathfrak{M}_6\,,
}
and the mass derivatives of the partition function was computed to all orders in $1/c_T$ in \cite{Binder:2018yvd} for the interacting sector of $k=1$ ABJM:
\es{Z}{
\frac{1}{c_T^2}\frac{\partial \log Z}{\partial m_+^2 \partial m_-^2} \Big\vert_{m_\pm=0}=-\frac{\pi ^2}{64c_T}+\frac{5 \pi ^{4/3}}{4\ 6^{2/3}c_T^{5/3}}+\frac{-\frac{5}{12}-\frac{15\pi^2}{64}}{{c_T}^2}-\frac{8 (2 \pi )^{2/3}}{ {3}^{\frac43}
  c_T^{7/3}}+\frac{1456+225 \pi ^2}{36 (6 \pi )^{2/3}
  c_T^{8/3}}+\dots\,.
}
After imposing both localization constraints, we find that the coefficients of the four contact term ambiguities are nontrivially consistent with the values given by the inversion formula analytically continued to low spins:
\es{RR4final3}{
(\lambda^{R|R^4}_{(A,+)_0})^2&=-\frac{539754496 \cdot 2^{\frac13}}{45\cdot 3^{\frac23}\pi^{14/3}}-\frac{1073741824 \cdot 2^{\frac13}}{429\cdot 3^{\frac23}\pi^{8/3}}\,,\\
(\lambda^{R|R^4}_{(A,+)_2})^2&=-\frac{14645368717312 \cdot 2^{\frac13}}{91875\cdot 3^{\frac23}\pi^{14/3}}-\frac{1832611005595648 \cdot 2^{\frac13}}{56581525\cdot 3^{\frac23}\pi^{8/3}}\,,\\
(\lambda^{R|R^4}_{(A,2)_1})^2&=-\frac{335829532672 \cdot 2^{\frac13}}{315\cdot 3^{\frac23}\pi^{14/3}}-\frac{6356551598080 \cdot 2^{\frac13}}{29393\cdot 3^{\frac23}\pi^{8/3}}\,,\\
(\lambda^{R|R^4}_{(A,2)_3})^2&=-\frac{3269552980885504 \cdot 2^{\frac13}}{848925\cdot 3^{\frac23}\pi^{14/3}}-\frac{4912617952903168 \cdot 2^{\frac13}}{6292363\cdot 3^{\frac23}\pi^{8/3}}\,,\\
   \gamma_{2,0}^{R|R^4}&=\frac{51018899456 \cdot 2^{\frac13}}{15\cdot 3^{\frac23} \pi ^{14/3}}+\frac{2849843046400 \cdot 6^{\frac13}}{46189 \pi ^{8/3}}\,,\\
\gamma_{2,2}^{R|R^4}&=\frac{110802345984 \cdot 6^{\frac13}}{5 \pi ^{14/3}}+\frac{433714857246720 \cdot 6^{\frac13}}{96577 \pi ^{8/3}}\,,\\
\gamma_{2,4}^{R|R^4}&=\frac{4515696959488 \cdot 2^{\frac13}}{63\cdot 3^{\frac23} \pi ^{14/3}}+\frac{1893141053440 \cdot 6^{\frac13}}{391 \pi ^{8/3}}\,.\\
}
This supports the conjecture that all such 1-loop contact terms vanish also for the $k=1$ theory.

\section{Numerical bootstrap}
\label{numBoot}

We will now compare the OPE coefficients computed to all orders in $1/c_T$ in the previous section, to finite $c_T$ numerical bootstrap bounds. In \cite{Agmon:2019imm}, bounds were computed on CFT data that appears in correlators of $S_2$ and $S_3$. The advantage of this mixed system is that we can explicitly rule out free theories by disallowing the free multiplet in $S_2\times S_3$, the appearance of $S_3$ automatically rules out the pure AdS$_4$ theory as well as the $k=2$ theory, and we also have access to more short $(B,+)$ and $(B,2)$ multiplets as listed in Table \ref{compare} that can be computed using localization. 

\begin{figure}[]
\begin{center}
   \includegraphics[width=0.49\textwidth]{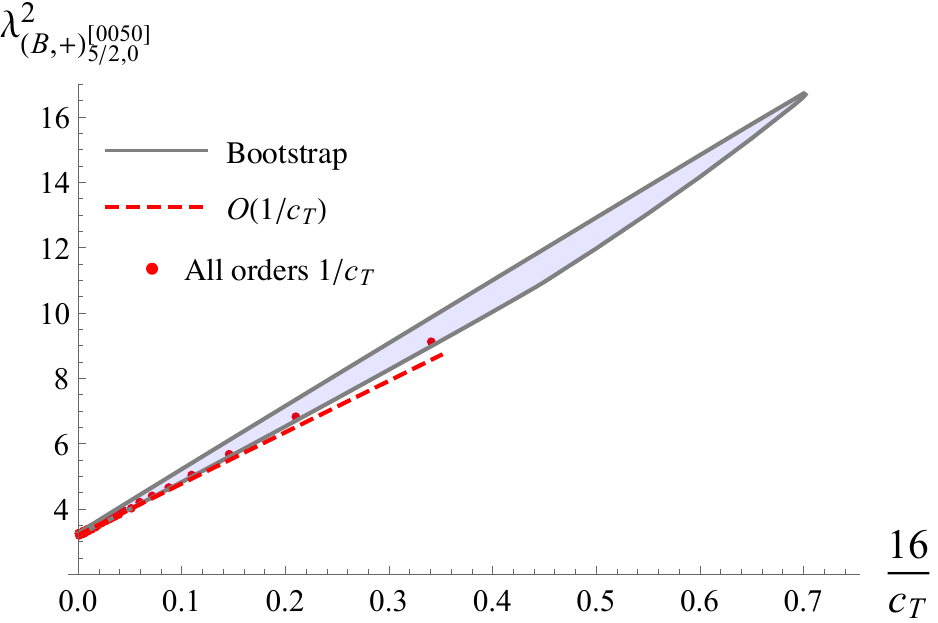}
      \includegraphics[width=0.49\textwidth]{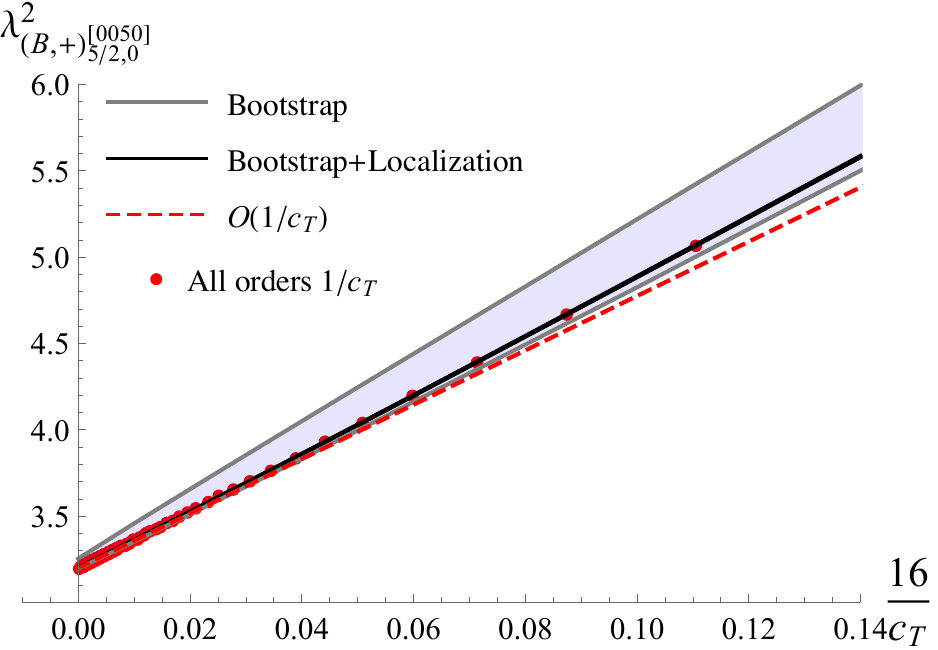}
 \caption{Upper and lower bounds on the ${(B,+)_{5/2,0}^{[0050]}}$ OPE coefficient squared in terms of the stress-tensor coefficient $c_T$ for the full range of $c_T$ ({\bf left}) and in the large $c_T$ regime ({\bf right}). The gray lines are bounds from mixed correlators of $S_2$ and $S_3$ with bootstrap precision $\Lambda=35$, while the black lines come from the same setup but with the OPE coefficients of all short multiplets except  ${(B,+)_{5/2,0}^{[0050]}}$ imposed using their all orders in $1/c_T$ expressions from localization. The red dotted line denotes the large $c_T$ expansion to order tree level supergravity $O(c_T^{-1})$, while the red dots denote the all orders in $1/c_T$ expression for integer values of $N$ starting with the interacting sector of $U(3)_1\times U(3)_{-1}$.}
\label{nobs}
\end{center}
\end{figure}

We start by comparing the most general bootstrap bounds for $\lambda^2_{(B,+)_{5/2,0}^{[0050]}}$ to the all orders in $1/c_T$ expression from localization. As shown in the lefthand plot of Figure \ref{nobs}, the localization values seem to approximately saturate the lower bound for the entire range of known theories, which extends from GFFT at $c_T\to\infty$ to the interacting sector of the $U(3)_1\times U(3)_{-1}$ ABJM theory.\footnote{The $U(2)_1\times U(2)_{-1}$ theory does not appear, because it does not have a $S_3$ operator.} Note that the all orders in $1/c_T$ expressions are noticeably different from the leading order $1/c_T$ expressions, which is the same for any theory with an Einstein gravity dual at large $c_T$. Also, we expect the all orders expression to be accurate even for the lowest $c_T$ theory of $U(3)_1\times U(3)_{-1}$ ABJM, as shown before in Table \ref{compare}.

We then focus on the large $c_T$ regime, to look for a more detailed comparison. As seen from the righthand plot of Figure \ref{nobs}, there is a slight discrepancy between the most general lower bootstrap bound and the all orders in $1/c_T$ expression. We can further strengthen the bootstrap bounds by imposing the values of all short OPE coefficients except $\lambda^2_{(B,+)_{5/2,0}^{[0050]}}$.\footnote{In more detail, since 1d crossing relates the OPE coefficients of the various short multiplets \cite{Agmon:2019imm}, we impose all the independent values except $\lambda^2_{(B,+)_{5/2,0}^{[0050]}}$, so these improved bounds truly come from 3d crossing, and are not trivially fixed by 1d crossing.} We now find that the upper/lower bounds collapse into a thin curve that exactly matches the all orders in $1/c_T$ expression. This supports the conjecture from \cite{Alday:2021ymb} that imposing the value of short OPE coefficients from localization is sufficient to fix the bootstrap bounds so that they correspond to the physical theory.

\section{Conclusion}
\label{conc}

This paper includes two main results. The first is the 1-loop $\langle2222\rangle$ Mellin amplitude for $k=1$ ABJM theory for $R$ and $R^4$ vertices, which generalizes previous results for $k=2$. This calculation presented new challenges relative to previous cases, as we needed to compute odd twist CFT data from tree level correlators whose position space expression could not be written in closed form. The second result is the explicit localization expression for protected OPE coefficients in $\langle22pp\rangle$ to the first few orders in large $c_T$, which we used to fix the contact term ambiguity for $p=2$, and compared to the numerical bootstrap for $p=3$. We found that after we impose one of the protected OPE coefficients, the bootstrap bounds for the other OPE coeffficients were saturated for all $c_T$ by the known expression, which suggests that these bounds can be used to numerically solve the $k=1$ ABJM correlator.

 Looking ahead, it would be nice to fix all the ambiguities in the $k=1,2$ $R|R$ 1-loop amplitudes, which are in principle fixed by the superconformal Ward identity, but are hard to fix in practice. While we can already extract any CFT data we want from the explicit position space double discontinuity, the full Mellin space expression would still be useful for integrated constraints such as \cite{Binder:2018yvd}. This integrated constraint will be essential to fix higher derivative terms in future studies. We would also like to understand better why the even and odd twist contributions to the Mellin amplitude are so similar. While at leading large $s,t$ this was required to get the right flat space limit, at subleading order this is more mysterious and could perhaps be explained by purely CFT arguments.
 
For the comparison to numerical bootstrap, we would like to compute 1-loop CFT data for the mixed correlators system, so we can compare them to the large $c_T$ regime of the bootstrap bounds for $k=1$ ABJM, as was successfully done in \cite{Alday:2021ymb} for the $k=2$ theory and the localization improved single correlator bounds. 
 
For the localization calculation, we would also like to compute non-perturbative corrections to the all orders in $1/N$ expression for the OPE coefficients, as was done for the ABJM sphere free energy in \cite{Marino:2009jd,Nosaka:2015iiw,Drukker:2010nc,Marino:2011eh,Hatsuda:2012dt,Hatsuda:2013gj,Calvo:2012du,Hatsuda:2013oxa,Kallen:2013qla,Honda:2013pea,Matsumoto:2013nya,Kallen:2014lsa,Codesido:2014oua}. Lastly, it would be nice to compute additional supersymmetric localization constraints from the squashed sphere, as was initiated in \cite{Chester:2021gdw,Chester:2020jay,Bomans:2021ldw,Minahan:2021pfv}.

\newpage
\section*{Acknowledgments} 

We thank Ofer Aharony, Davide Gaiotto, Silvu Pufu, and Xinan Zhou for useful conversations. The work of LFA is supported by the European Research Council (ERC) under the European Union's Horizon
2020 research and innovation programme (grant agreement No 787185). SMC is supported by the Weizmann Senior Postdoctoral Fellowship. The work of LFA is supported by the European Research Council (ERC) under the European Union's Horizon 2020 research and innovation programme (grant agreement No 787185).  LFA is also supported in part by the STFC grant ST/T000864/1. The authors would like to acknowledge the use of the University of Oxford Advanced Research Computing (ARC) facility in carrying out this work.(http://dx.doi.org/10.5281/zenodo.22558)
\appendix

\bibliographystyle{JHEP}
\bibliography{3ddraft}

\end{document}